\documentclass[aps,twocolumn]{revtex4}
\usepackage{graphicx}
\begin{document}

\title{Planar nanocontacts with atomically controlled separation}
\author{Y.V. Kervennic, L.P Kouwenhoven, and H.S.J Van der Zant}
\affiliation{Kavli Institute of Nanoscience, Faculty of Applied
Sciences, Delft University of Technology, Lorentzweg 1, 2628 CJ
Delft, the Netherlands}
\author {D. Vanmaekelbergh}
\affiliation{Chemistry and Physics of Condensed Matter, Debye Institute, University of Utrecht, Princetonplein 1, 3584 CC Utrecht, The Netherlands }

\begin{abstract}

We have developed a technology to reproducibly make gaps with distance control on the single atom scale. The gold contacts are flat on the nanometre scale and are fabricated on an oxidized aluminium film that serves as a gate. We show that these contacts are clean and can be stabilized via chemical functionalization. Deposition of conjugated molecules leads to an increase in the gap conductance of several orders of magnitude. Stable current-voltage characteristics at room temperature are slightly nonlinear. At low temperature, they are highly nonlinear and show a clear gate effect.

\end{abstract}

\maketitle

An important problem in molecular electronics consists of fabricating three-terminal devices with nano-objects. Several methods have been developed~\cite{review}. In the mechanical approach (STM and break junctions \cite{hydrogen,reed}) piezoelectric control of the gap distance offers great precision well below the atomic scale. Transport through small molecules have been measured \cite{hydrogen,reed2,kergueris,Reichert}, but implementation of a gate is difficult. Electromigration~\cite{park} of gold wires on top of aluminium or silicon gate is in that respect more promising and has shown impressive results \cite{pseudoatomset,pseudopark}. Nevertheless the electromigration method only provides statistical control with a low yield and no control on the atomic scale. Electrochemical deposition techniques \cite{Tao,Morpurgo,Ker2002} enable in principle control of nanogaps on the atomic scale with a high yield, a necessary requirement for integration of molecular devices in circuits. A disadvantage of electrochemical deposition is the formation of relatively thick electrodes. Gating is then a problem if the smallest distance is far from the substrate.

In this letter we show that flat, stable contacts with a separation ranging from one atom up to a few nanometers can be made on aluminium gates. We use electrochemical etching of very thin gold films ($\sim$15 nm) while monitoring the conductance of the gap. Subatomic resolution is obtained with a high yield. After rinsing in demineralized water, the contacts are clean enough to be directly functionalized with molecules containing thiol or sulfide groups. Measurements of current-voltage characteristics show that they then form stable contacts for nano-objects.

\begin{figure}[htbp]
 % \begin{center}
\centering
 \vspace{-0cm}
\includegraphics[width=9cm]{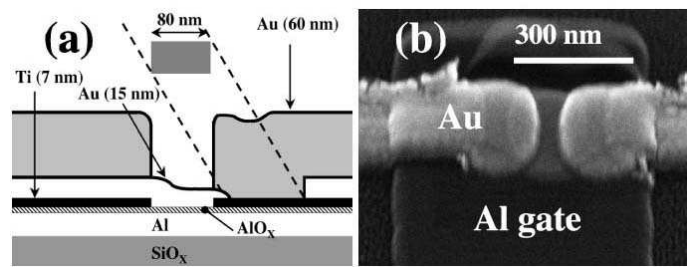}
  \vspace{0cm}
    \caption{Device configuration. (a) Fabrication process by shadow evaporation. The thinner central part is obtained by small-angle evaporation (15$^{\rm o}$ from vertical) of gold in the shadow of the resist bridge. (b) A SEM micrograph of a thin gold link sitting directly on aluminium oxide. Two thick gold electrodes connect the thin Au film (link) to large pads that are used for contacting.}
    \label{figc1}
  \vspace{0cm}
\end{figure}

Fabrication consists of two major steps: evaporation of a thin gold link on top of an oxidized aluminum film and, secondly, electrochemical etching of the gold to form a nanogap. In the first major step, we start with evaporating a 20~nm thick, 600~nm wide aluminum wire on an oxidized undoped silicon substrate (see also Fig$.$ \ref{figc1}). On top, we define two lines (the electrodes) separated by about 100~nm using e-beam lithography in a two-layer resist. After resist development, the evaporation sequence is as follows: first a 7~nm sticking layer of titanium is evaporated perpendicularly. The sample is then tilted by 15$^{\rm o}$ and the 15~nm thin gold link is formed. Finally, the thick gold pads are deposited at 90$^{\rm o}$.

%%%%%%%%%%%%%%%%%%%%%%%%%%%%%%%%%%%%%%%%%%%%%%%%%%%%%%%%%%%%%%%%%%%%%%%%%%%%%%%%%%%%%%%%%%%%%%%%%%%
%figure1
%%%%%%%%%%%%%%%%%%%%%%%%%%%%%%%%%%%%%%%%%%%%%%%%%%%%%%%%%%%%%%%%%%%%%%%%%%%%%%%%%%%%%%%%%%%%%%%%%%%

In the second major step, we slowly reduce the weak link thickness using electrochemical etching in a bath of KAuCN$_{2}$ (0.1~mM of KAuCN$_{2}$, 0.2~M of KOH and 1~M of KHCO$_{3}$). As counter electrode, we use a small piece of silicon ($\sim 20$~mm$^2$) covered with 100~nm evaporated gold. The voltage on this counter electrode ($V_{\rm CE}$) sets the etching rate. The electrodes are contacted using spring contacts, and only a small part of the gold electrodes ($<0.05$~mm$^2$) is dipped in the etching bath. During the etching process, the inter-electrode conductance is measured using an $I$-$V$ converter and a lock-in amplifier to detect the small AC responses. Gap resistances can be measured up to 1~G$\Omega$. Note that from comparing the gap resistance in solution and after drying, we conclude that conduction enhancement by the solution is not significant.

\begin{figure}[htbp]
 % \begin{center}
\centering
\includegraphics[angle=0,width=8.5cm]{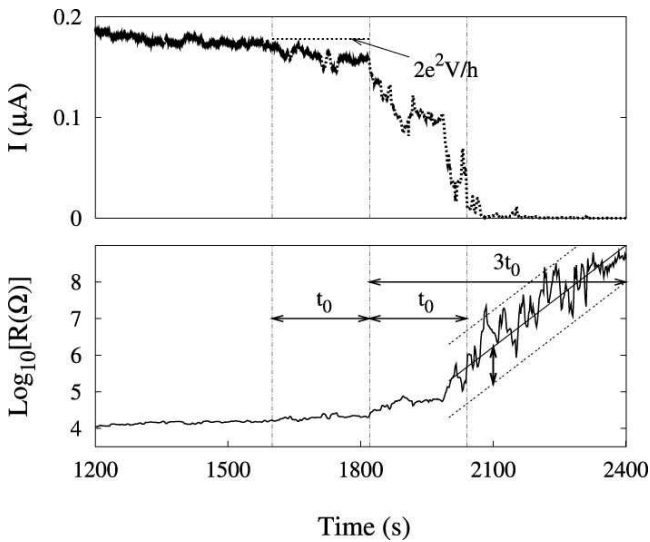}
% {\epsfig{file=figure1.jpg,angle=0,width=9cm, clip=true}}
\vspace{0cm}
    \caption{Gap current (upper graph) and resistance (lower graph) versus etching time. Only the slow etching part, in which the gap is formed, is shown. $V_{\rm CE}$ is kept constant at - 1470~mV. The excitation bias was 10~mV with a frequency of 5~Hz. For $t>2000$~s, the gap resistance shows an approximately exponential time dependence. The dashed lines correspond to a resistance deviation equivalent to about 0.5 atomic length. Time indication ($t_{0}$) has been added to characterize the last plateau length and the gap opening. The current value $\huge\frac{2e^2}{h}$V indicates the conductance quantum.}
    \label{figc2}
 % \end{center}
  \vspace{0cm}
\end{figure}

The electrochemical etching proceeds in two parts. We start etching the gold at a high current density with $V_{\rm CE} = -2400$~mV. The excitation frequency of the lock-in is set to 41~Hz \cite{nota1} and the measured signal is fed through a high pass filter to remove the large DC etching current. After a few minutes, the inter-electrode conductance starts to drop significantly. When the resistance reaches a value of about 3~k$\Omega$, $V_{\rm CE}$ is changed to -1560~mV. The etching rate is now so slow that removal of a single atom from the gap region can be observed. Fig$.$ \ref{figc2} shows one out of the 100 very similar traces of the current through the gap vs. time. We find that when the conductance is around G$_{0}$ = ${\frac{2e^2}{h}}$, the current trace becomes more noisy for a time span of about $t_{\rm 0}$. A sharp decrease in the current  follows and the gap current starts to show large fluctuations. These observations suggest that $t_{\rm 0}$ is the typical time scale to remove the last gold atom from the contact. The noise that occurs after opening of the gap, remains after washing and drying. We attribute it to gold-atom motion on the surface. More details of the fabrication process and the stability will be published elsewhere \cite{Ker2003}.

%%%%%%%%%%%%%%%%%%%%%%%%%%%%%%%%%%%%%%%%%%%%%%%%%%%%%%%%%%%%%%%%%%%%%%%%%%%%%%%%%%%%%%%%%%%%%%%%%%%
%%%%%%%%%%%%%%%%%%%%%%%%%%%%%%%%%%%%%%%%%%%%%%%%%%%%%%%%%%%%%%%%%%%%%%%%%%%%%%%%%%%%%%%%%%%%%%%%%%%
An interesting feature of our data is that the etching rate is reproducible even at the atomic scale. As shown in Fig.~\ref{figc2} the gap resistance increases close to exponential with etching time after opening of the gap. Since tunneling is exponential in distance, this indicates an approximate constant etching rate on the atomic scale. Moreover for all our samples, we find that the time span to open the gap to 1~G$\Omega$ equals $\sim$$3~t_{\rm 0}$ when starting from the sudden decrease of the current. A gap with a resistance 1~G$\Omega$ is estimated~\cite{reed,park} to be 3 atoms wide ($\sim$ 1nm) in agreement with our previous estimate that $t_{\rm 0}$ is the time scale to remove one gold atom. We also find that one time $t_{\rm 0}$ after the last atom removal, the resistance is always close to 1~M$\Omega$. This value then corresponds to a gap of one atom wide. Furthermore, as illustrated in Fig.~\ref{figc2}, the noise has an amplitude that corresponds to the motion of gold atoms over half an atomic distance ($\sim 1.5$~\AA), indicating that atomic resolution is achieved.

After rinsing in water and drying the gaps, we have checked the gate at room temperature. We generally find very small leakage currents (smaller than 10~pA at 1~V for more than 50~\% of the samples). The typical gate voltage beyond which current starts to grow significantly is $\pm 2$~V. Importantly, scanning electron microscope (SEM) imaging also indicates that the aluminium gates are not affected by the etching process.

We have studied the morphology of the thin etched electrodes with a SEM. Near the gap, we find grains that vary in size between 15~nm and 8~nm. Smaller grains are not observed even at high resolution. The gap region is generally barely visible indicating an extreme shallowness compared to other remaining parts of the thin gold link (with atomic force microscopy we were also unable to resolve the height of the end parts of the electrodes). This clearly indicates that the geometry is planar and that the point of smallest gap distance is directly on top of the aluminium gate. The use of aluminium is crucial: thin gold films on silicon oxide break during etching before atomic resolution is obtained. Experimentally and theoretically it has been shown that in ultra-thin films on aluminium oxide, gold atoms strongly bind to the negatively charged oxygen sites, which may account for the greater stability \cite{TEM,Altheory}.

%%%%%%%%%%%%%%%%%%%%%%%%%%%%%%%%%%%%%%%%%%%%%%%%%%%%%%%%%%%%%%%%%%%%%%%%%%%%%%%%%%%%%%%%%%%%%%%%%%%%%%%%%%%%%%%%%%%%%%%%%
\begin{figure}[htbp]
\centering
 % \begin{center}
%\hspace{-10cm}
%\vspace{-0.5cm}
\includegraphics[angle=0,width=7cm]{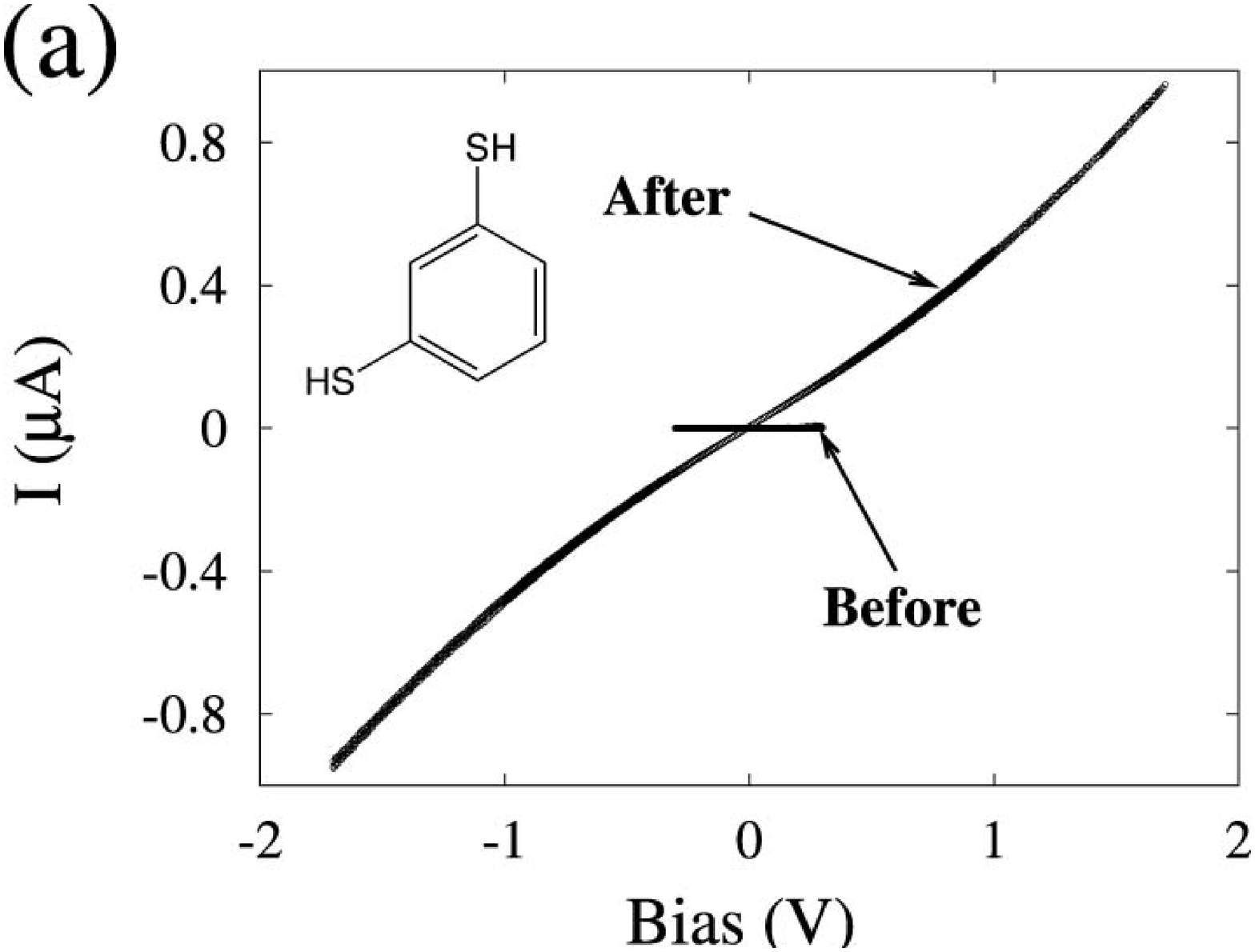}
\includegraphics[angle=0,width=7cm]{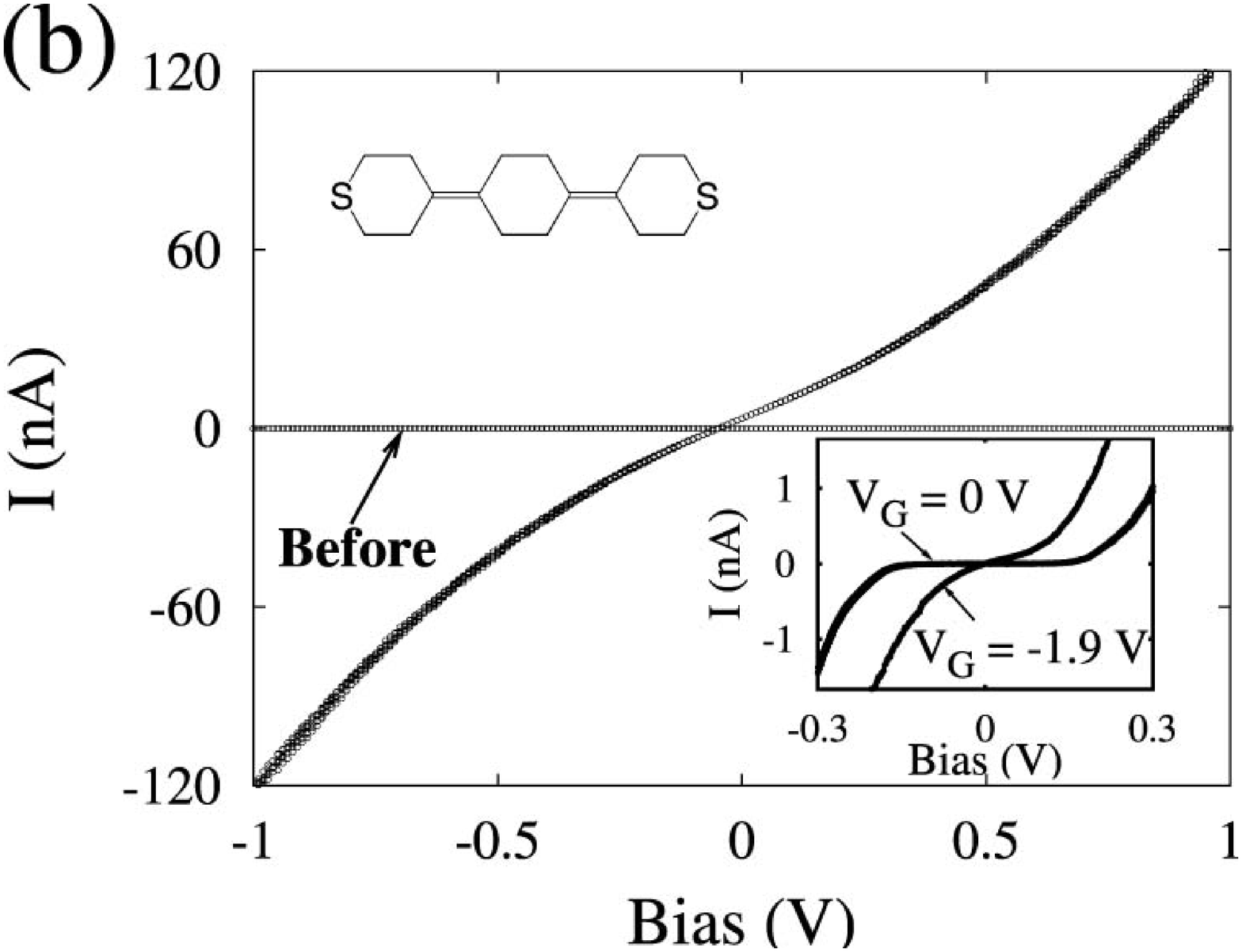}
 %\centerline{\epsfig{file=figure3.jpg,angle=0,width=5cm, clip=true}}
    %\end{minipage}\hfill
    %\begin{minipage}[c]{0.48\linewidth}
      \caption{(a) Room temperature $I-V$ characteristic for a 4 atom gap covered with 1-3 disulfobenzene. The increase in current is of more than two orders of magnitude. (b) Characteristics for a gap covered with oligo(cyclohexylidenes). Inset : $I-V$ curves at 4.2 K for two values of the gate voltage. As can clearly be seen, the gap of $\sim 300$~mV can completely be suppressed. For all samples the deposition time was 24 hours.}
  \label{figc3}
\end{figure} 
%%%%%%%%%%%%%%%%%%%%%%%%%%%%%%%%%%%%%%%%%%%%%%%%%%%%%%%%%%%%%%%%%%%%%%%%%%%%%%%%%%%%%%%%%%%%%%%%%%%

%\begin{figure}[htbp]
  %\begin{center}
   %\centerline{\epsfig{file=./fig/figure3.pdf,angle=0,width=14cm, clip=true}}
   % \caption{SEM micrographs of three electrode pairs of different separation with individual Au clusters trapped in between %them. From left to right the cluster size is 20~nm, 10~nm and 5~nm. No cleaning prior to SEM imaging is possible without %damaging the sample. This explains why the images of these samples are less sharp than those shown in %Fig$.$~\ref{figch41}.}
    %\label{figch3}
  %\end{center}
  %\vspace{0.5cm}
%\end{figure}

By measuring the gap resistance, we find that bare gaps are stable on the scale of about one hour in ambient conditions. Stability is drastically improved by functionalization of the surface by molecules with thiol or sulfide ends. As a test that the gaps can be used for transport measurements of single or a small number of molecules, we have dipped them in millimolar ethanolic solutions of 1-3 benzene dithiol and oligo(cyclohexylidene)-disulfides. Such molecules are rigid, have $\pi$ or $\sigma$ conjugation in the backbone and have gold-compatible thiol or sulphide end functions~\cite{molecules,molecules2}. In all cases a clear increase of conductance is measured after deposition of the molecules. Current-voltage characteristics are slightly nonlinear at room temperature as illustrated in Fig$.$ \ref{figc3}. 
In the case of oligo(cyclohexylidenes), bare gaps have initial resistances that are too high to be measured in our set-up ($I<1$~pA for $V=1$~V, see Fig$.$~\ref{figc3}). The length of this molecule is 1.18 nm, a distance comparable to that of 4 gold atoms in a row. After dipping in the solution with the molecules, gaps of about 4 atoms wide (to be more precise corresponding to 4-5~$t_0$) show a resistance of $\sim$ 1 M$\Omega$. For slightly larger gaps corresponding to 6-8~$t_0$, resistances of the order of 10~M$\Omega$ are reproducibly found.

Several experiments have been performed to show that transport through the molecules is responsible for the strongly enhanced current. First, we have dipped over 10 gaps in ethanol for more than 24 hours and did not find any decrease of the resistance. In addition, for gaps that are larger than 2.5 molecular length -so that two molecules facing each other but attached to different electrodes do not touch-, the resistance after dipping in the oligo(cyclo hexylidenes) solution is orders of magnitude larger (i.e., $>1$~G$\Omega$).

$I$-$V$ traces have been measured over period of ten hours at room temperature and we find that the stability is much better than bare gaps. Occasionally, sudden switchings occur, indicating reconfigurations of the gold atoms contacting the molecules. Furthermore, when kept in vacuum the resistance of the molecular nanodevice remains approximately the same (over the time span of weeks).

Preliminary measurements have been performed at low temperatures. $I$-$V$ characteristics show gap-like curves as illustrated in the inset of Fig$.$ \ref{figc3}. For the four sample studied, the same gap of $\sim 300$~mV was found; for the three samples with the gate connected, a clear gate effect has been observed (see inset of Fig$.$ \ref{figc3}). Interestingly, the gate effect presents unconventional features (no saturation, nor periodicity as observed in coulomb blockade). A systematic study of the gate effect and the low temperature behavior is beyond the scope of this paper and will be presented elsewhere.

In conclusion, we have made flat gold nanogaps with atomic precision. The fabrication process is compatible with aluminium gating technology and yields gaps that can readily be functionalized with standard dithiol linkers. Once bound chemically to rigid nano-objects, these gaps are stable for weeks. Our method is compatible with current on-chip technology and provides a simple, fast and reproducible technique for the fabrication of molecular and few-atoms based nanodevices.
\\
%\clearpage
\indent
We are thankful to R.M. Schouten for technical advice concerning electronics and L. W. Jenneskens for the provision of the oligo(cyclohexylidenes). We also thank Alberto Morpurgo, Richard Deblock, Gilles Gaudin, Stefan Oberholzer, Monica Monteza, Sebastien Viale for discussions, and Bert de Groot for technical assistance. This work was supported by the Dutch Organization for Fundamental Research on Matter (FOM).

\vspace{0.5cm}

\noindent 
%  \bibliography{apssamp}% Produces the bibliography via BibTeX.

\begin{thebibliography}{99}
\bibitem{nota1} To perform the maximum resolution ($\sim$ pA), the frequency has to be reduced to a few hertz at low etching current to minimize the influence of parasitic capacitances.
\end{thebibliography}

\end{document}